\documentclass[twocolumn,prl,aps,epsfig,floats,showpacs]{revtex4}
\def\DESepsf(#1 width #2){\epsfxsize=#2 \epsfbox{#1}}
\usepackage{epsfig}
\usepackage{graphicx}
\def\bmatrix{\left[\begin{array}}
\def\ematrix{\end{array}\right]}

\begin{document}
\title{\boldmath
Resolving the $B\to \phi K^*$ Polarization Anomaly }
\vfill
\author{
Wei-Shu Hou and Makiko Nagashima }
\affiliation{ \rm Department of Physics, National Taiwan
University, Taipei, Taiwan 106, R.O.C. }

%
%
\vfill
\begin{abstract}
The experimental observation of sizable transverse components for
$B\to \phi K^*$ decay is in strong contrast to all other $VV$
modes, and poses a challenge to our understanding of $B$ decay
dynamics.
Observing that the gluon emitted from $b\to sg^{(*)}$ chromodipole
transition is transverse, we give a heuristic model where the
transverse $\phi$ descends from the emitted gluon, hence similar
phenomena should occur for $B\to \omega K^*$ but not for $B\to
\rho^0 K^*$.
New physics in $bsg$ chromodipole coupling, perhaps needed for the
$\bar B \to \phi K_S$ $CP$ violation anomaly, may lead to
different patterns of $CP$ and $T$ violation in transverse
components of $\bar B \to \phi \bar K^*$, $\omega \bar K^*$
decays.
\end{abstract}
\pacs{PACS numbers:
11.30.Hv, 
12.60.Jv, 
11.30.Er, 
13.25.Hw  
}
%
\maketitle

\pagestyle{plain}

Two-body charmless $B\to PP$, $PV$ and $VV$ decays, where $P$, $V$
stand for light pseudoscalar and vector mesons, can provide us
access to quark mixing and $CP$ violation parameters, and insight
into strong dynamics. The $VV$ modes are more difficult to study
as they involve $s$-, $p$- and $d$-wave components. Some modes
have appeared recently~\cite{PDG,HFAG}, bringing forth, however,
the so-called $B\to \phi K^*$ polarization anomaly. Both
Belle~\cite{Belle03} and BaBar~\cite{BaBar03,BaBar04} experiments
have observed significant transverse components of $B\to \phi K^*$
decay, while theoretically it is argued~\cite{Kagan} that these
should be $1/m_b^2$ suppressed. We do not know whether it is
related to the ``$CP$ violation anomaly" in $\bar B \to \phi K_S$,
but since Belle~\cite{phiKsBelle} and BaBar~\cite{phiK0BaBar} are
at variance on the latter, with BaBar in agreement with Standard
Model (SM) expectations, the $\phi K^*$ polarization anomaly may
be viewed as more urgent. So far there are no convincing
solutions.

In this paper we offer a possible solution. We note that on-shell
$b\to sg$ decay has a rate of a few $\times 10^{-3}$~\cite{Hou88},
and the emitted gluon is dominantly transverse. Viewing the
transverse $\phi$ meson ($\phi_T$) as a leading single particle
gluon fragment, it is conceivable that $B\to \phi_T K^*_T$ at few
$\times 10^{-6}$ can be generated. The feeddown fraction is two
orders of magnitude smaller than the $B\to K^*\gamma$ case, which
is about 13\% of $b\to s\gamma$ rate~\cite{PDG}. It follows that
one should find transverse component for $B\to \omega K^*$ mode
but not for $\rho K^*$, $\rho\rho$. The mechanism also opens a
window onto possible New Physics (NP) in chromodipole $bsg$
coupling, which may have already manifested itself in the $\phi
K_S$ $CP$ violation anomaly.

The $1/m_b$ suppression of transverse component can be seen
heuristically as follows. The longitudinal polarization for a
vector meson $V$ can be approximated by $\epsilon_L^\mu \to
p^\mu/m_V$ up to $O(m_V/E)$. Since $E_{1,2}\sim M_B/2$ for
charmless $B\to V_1V_2$ modes, we see that $\epsilon_{1L} \cdot
\epsilon_{2L} \to p_1\cdot p_2/m_1m_2$ is of order $M_B^2$,
whereas $\epsilon_{1T} \cdot \epsilon_{2T}=-1$. The dominance of
longitudinal component seems to be borne out by $B\to
\rho\rho$~\cite{rrBelle,rrBaBar} and $\rho K^*$~\cite{BaBar03},
but it apparently breaks down for $B\to \phi K^*$. In the linear
polarization ($CP$) basis, the experimental results
are~\cite{Belle03,BaBar03,BaBar04}
\begin{eqnarray}
& & |f_0|^2 = 0.43 \pm 0.09 \pm 0.04 \hspace{5mm} {\rm (Belle)},
\nonumber \\
& & |f_0|^2 = 0.52 \pm 0.05 \pm 0.02 \hspace{5mm} {\rm (BaBar)},
\end{eqnarray}
for the longitudinal fraction, and
\begin{eqnarray}
& & |f_\perp|^2 = 0.41 \pm 0.10 \pm 0.04 \hspace{5mm} {\rm
(Belle)},
\nonumber \\
& & |f_\perp|^2 = 0.22 \pm 0.05 \pm 0.02 \hspace{5mm} {\rm
(BaBar)},
\end{eqnarray}
for the CP-odd fraction.

It has been argued~\cite{Kagan} that the chromodipole $b\to s\bar
ss$ 4-quark operator $O_{12}$ cannot contribute to transverse
$\phi$ formation. This is because the $s$ quark from $b\to s$
dipole transition is paired with the $\bar s$ quark from the
virtual gluon. Since the latter is transverse, one would always
have a mismatch of quark helicities.
Our proposal also starts from the transverse nature of such a
gluon, but we focus on the case where it is close to the mass
shell. Being rather energetic, such a gluon reaches the $B$ meson
surface in less than $10^{-24}$~s without much interaction
(perturbative vacuum), but then it must shed its color. Its
``essence" should be able to penetrate the meson surface at ease,
i.e. the energy, momentum and perhaps its angular momentum would
depart from the $B$ meson carcass ``instantly", leaving behind
some hadronic scale disturbance to balance the color.

\begin{figure}[b!]
\centerline{\hskip0.05cm\epsfxsize2.1in \epsffile{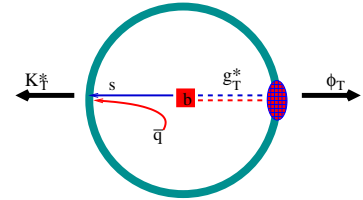}}
%
\caption{ Heuristic picture for transverse $\phi$ emission. The
gluon from $b\to s$ chromodipole is mostly transverse, and could
emit a transverse $\phi$ meson. The singlet nature of the gluon
implies that this process does not affect charged vector meson or
$\rho^0$. }
 \label{fig:heuristic}
\end{figure}

The above heuristic picture is depicted in Fig. 1. The system
recoiling against $\phi_T$ consists of the fast $s$ quark, the
spectator $\bar q$, and the color octet remnant (the minimum would
be two soft gluons) of the gluon shown as the ellipse, which is
rather complicated. But if it ends up as a $K^*$, it would also be
transverse by angular momentum conservation.

Let us start from the relevant effective Hamiltonian
\begin{eqnarray}
H_{\rm eff} &=& - \frac{G_f}{\sqrt 2}V_{ts}^* V_{tb} \Bigl\{
\sum_{n=3-10}c_n\,O_n
\nonumber \\
& & + c_{12} \, \frac{g_s}{8\pi^2}m_b \bar s_i \sigma_{\mu\nu}
(1+\gamma_5) T^a_{ij} G^{a\mu\nu} b_j \Bigr\},
 \label{eq:gqq}
\end{eqnarray}
where $n = $3--6 and 7--10 label strong and electroweak penguins,
respectively, and the chromodipole operator is explicitly
dimension 5. While $O_{3-10}$ are local, the 4-quark operator
$O_{12}$ generated by the chromodipole term is non-local,
indicating its special character. After all, the gluon from the
parton level $b\to sg$ process can propagate a long distance,
giving rise to an inclusive rate $\sim (2-3)\times 10^{-3}$ in SM.
{\it The process is the QCD analog of the famous $b\to s\gamma$
transition}, but its effect has been rather elusive
experimentally, while theorists tend to treat it as an
afterthought within the operator framework.

The ``on-shell" gluon is dominantly transverse. Taking $q^2 \sim
1$ GeV$^2$ as an effective gluon mass, such an energetic ($E\sim
m_b/2$) gluon traversing the hadronic medium cannot be
distinguished from a colinear color octet $q\bar q$ pair. In any
case, it traverses the hadronic sized $B$ meson carcass in $\sim
10^{-24}$ s, much shorter than the hadronic time scale of $\sim
10^{-23}$~s, hence has little time to change its nature. When it
reaches the meson boundary, although the confinement energy would
quickly rise, so long the effective color octet $q\bar q$ pair
leaves behind the same color octet charge to settle its ``debt",
it can depart the $B$ meson carcass in a color singlet
configuration and hadronize. The remaining hadronic scale color
octet ``disturbance" (minimum of two soft gluons) balances the
color octet fast $s$ quark plus spectator $\bar q$ quark, and the
system takes the hadronic time scale of $\sim 10^{-23}$ s to
settle into a particular hadronic configuration, with some
amplitude as a single $K^*$ meson. Thus, our picture is the
combined effect of (transverse) gluon fragmentation plus recoil
side recombination.

Our argument is not perturbative, but boils down to an Ansatz of
replacing $T^a_{ij} G^{a\mu\nu}$ first by (ignoring constant
factors) $T^a_{ij} q^\mu\varepsilon_{g^*}^{*a\nu}$, then by
$\delta_{ij}p_\phi^\mu\varepsilon_{\phi}^{*\nu}$ where
$p_\phi^\mu$ is the $\phi$ momentum and differs from $q^\mu$ only
by a hadronic scale momentum.
%
%
Finally, we parametrize this mechanism of $\phi_T$ generation by a
hadronization parameter $\kappa$,
\begin{eqnarray}
\frac{\kappa}{m_B} \, c_{12} \, p_\phi^\nu \epsilon^{* \mu}_{\phi
T} \langle K^*_T|\bar s_i i\sigma_{\mu\nu} (1+\gamma_5) b_j | B
\rangle,
 \label{eq:Ansatz}
\end{eqnarray}
together with a factor $- \frac{G_F}{\sqrt 2} V_{ts}^\ast V_{tb}
f_\phi m_\phi$. The $f_\phi m_\phi$ factor is to conform with the
operators $O_{3-10}$ which produce $\phi$ from a vector current.
The rather complicated perturbative and nonperturbative
hadronization has been simplified into a single parameter, which
we take as real since there is no clear physical cut. Note that we
have absorbed $g_sm_b$ etc. into $\kappa$. To keep $\kappa$
dimensionless, the additional power of $1/m_B$ is in anticipation
of a similar factor from the hadronic matrix element.

We now write down the longitudinal (0), parallel ($\parallel$) and
perpendicular ($\perp$) amplitudes for $B\to\phi K^*$,
\begin{eqnarray}
{\cal A}_{i,\perp} &\propto&
 \Bigl\{
  \Bigl[\sum_{j=3,4,5}(a_j \mp a^\prime_j)
   - \frac{1}{2}\sum_{j=7,9,10}(a_j \mp a^\prime_j)
 \Bigr]X_\lambda
\nonumber \\
& & + (c_{12} \mp c_{12}^\prime)
 \Bigl[ \kappa_\lambda \tilde F_\lambda
  + \frac{\alpha_s}{4\pi}\frac{m_b^2}{q^2}\tilde S_\lambda \Bigr]\Bigr\},
 \label{eq:Amp}
\end{eqnarray}
in linear polarization basis, where $\lambda = i, \perp$ with $i =
0, \parallel$.
Epitomizing our Ansatz which does not feed the longitudinal mode,
$\kappa_0 \approx 0$, while $\kappa_\parallel \approx \kappa_\perp
= \kappa$.
The hadronic parameters
\begin{eqnarray}
& & X_0= \frac{m_B+m_K^\ast}{2}A_1\;x
        -\frac{m_\phi m_{K^\ast}}{m_B+m_{K^\ast}}A_2\;(x^2-1),
\nonumber \\
& & X_\parallel= -\frac{m_B+m_K^\ast}{\sqrt 2}A_1,
\nonumber \\
& & X_\perp = -\frac{m_\phi m_{K^\ast}}{m_B+m_{K^\ast}}V\;
\sqrt{2(x^2-1)},
\end{eqnarray}
are $B\to K^*$ form factor combinations. Noticing that
$x=p_\phi\cdot p_{K^\ast}/m_\phi m_{K^\ast}$ is of order $m_b^2$,
one can already see the $1/m_b$ suppression at work when one
compares $X_{\parallel,\perp}$ with $X_0$.
The usual chromodipole hadronic parameters $\tilde S_\lambda$ are
form factor combinations analogous to $X_\lambda$, and arise from
the $B\to K^*$ dipole in the matrix element of $O_{12}$. Again,
$\tilde S_0$ is larger than the other two.
Our model gives additional hadronic parameters $\tilde
F_\parallel$ and $\tilde F_\perp$ from Eq.~(\ref{eq:Ansatz}),
which are roughly
$X_\parallel$ and $-X_\perp$, but with dipole form factors.
Note that in Eq.~(\ref{eq:Amp}) we have kept opposite chirality
operators which are vanishingly small in SM, but may arise from
NP.

%
%
%
\begin{figure}[t!]
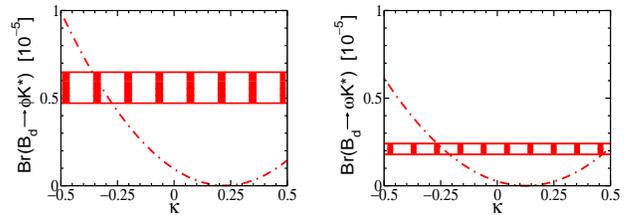

\smallskip  
\includegraphics[width=1.5in,height=1.1in,angle=0]{phiKstar_sm}
\hspace{2.7mm}
\includegraphics[width=1.5in,height=1.1in,angle=0]{omegaKstar_sm}
\smallskip\smallskip
\caption{ (a) ${\cal B}(B\to \phi K^*)$ and (b) ${\cal
B}(B\to\omega K^*)$ vs hadronization parameter $\kappa$ in
longitudinal (solid) and transverse (dotdash) components.
 }
 \label{fig:brSM}
\end{figure}

The branching fractions for $B\to \phi K^*$ in different
polarization components are plotted in Fig. 2(a). The longitudinal
component is independent of our new hadronic parameter $\kappa$,
but receives some suppression from interference with usual
chromodipole term. We have computed the coefficients $a_j$ and
$c_{12}$ at $\mu = m_b$ scale, which, together with light-cone sum
rule form factors~\cite{LCSR}, give longitudinal rate consistent
with data. We plot the sum of parallel and perpendicular
components since both are transverse and are of similar strength
in our Ansatz. With our somewhat {\it ad hoc} term, the transverse
component rises to $5 \times 10^{-6}$ for $\kappa \sim -0.25$, and
is enhanced slightly by usual chromodipole term.
%
The longitudinal rate can be further reduced if one uses different
form factors
where $A_2/A_1$ is larger, making $X_0$ smaller. Alternatively,
there is much uncertainty in the hadronic parameter $\tilde
S_0/q^2$. We illustrate the destructive interference and show, in
Fig.~2(a), the range of reduction when $q^2/m_b^2$ drops from 1/3
to 1/4.
It should be clear that there is no need for NP, and our Ansatz is
able to accommodate the $\phi K^*$ polarization anomaly within SM.

The Ansatz allows some immediate predictions.
The transverse gluon fragmentation picture should apply to
$\omega_T$ emission. Besides SU(3) breaking effects, one gains a
$\sqrt 2$ isospin factor.
The usual contributions to the decay amplitudes are now more
involved, where the tree contributions distinguish between charged
and neutral modes. Further, besides $\omega$ emission with $B\to
K^*$ transition terms, one now also has $K^*$ production with
$B\to \omega$ transition. Our Ansatz contributes clearly only to
the former.
We plot the $B\to \omega K^*$ results vs $\kappa$ (treated on same
footing as $\phi K^*$ case) in Fig.~2(b). We use $\phi_3 \equiv
\arg V_{ub}^* \simeq 60^\circ$ such that the charged and neutral
modes have very similar rates. The $B\to \omega K^*$ rate is
predicted to be of order $4\times 10^{-6}$, and the transverse
components could dominate.
Note that our mechanism would not feed $\rho^0 K^*$ channels, nor
$\rho^\pm K^*$. Thus, another consequence of our model is that the
$\rho K^*$ and $\rho\rho$ modes would be predominantly
longitudinal, in agreement with
data~\cite{BaBar03,rrBelle,rrBaBar}.

%
%
%
\begin{figure}[t!]
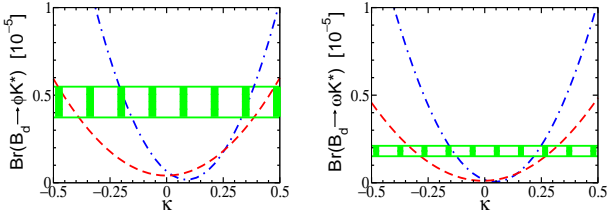

\smallskip  
\includegraphics[width=1.5in,height=1.1in,angle=0]{phiKstar_np}
\hspace{2mm}
\includegraphics[width=1.5in,height=1.1in,angle=0]{omegaKstar_np}
\smallskip\smallskip
\caption{ (a) ${\cal B}(B\to \phi K^*)$ and (b) ${\cal
B}(B\to\omega K^*)$ vs $\kappa$ for $\lambda = 0$ (solid),
$\parallel$ (dotdash) and $\perp$ (dash), with $CP$ phase $\sigma
= \pi/2$. }
 \label{fig:brNP}
\end{figure}

Although our model can be viewed as effective within SM, the $\phi
K_S$ $CP$ violation anomaly, if it persists, may call for the need
for NP. We have proposed~\cite{CHN} a NP model with a light,
flavor mixed ``strangebeauty" right-handed squark, the
$\widetilde{sb}_{1R}$, which brings in strong
$\widetilde{sb}_{1R}$-$\tilde g$ penguin loop, together with a new
$CP$ violation phase $\sigma$ from the $\tilde s_R$-$\tilde b_R$
squark sector. The model can be motivated~\cite{ACH} by
approximate Abelian flavor symmetry which implies large
right-handed $s$-$b$ mixing, transferred to the squark sector, and
drives the $\widetilde{sb}_{1R}$ light. It can be as light as
100--200 GeV with common squark mass at TeV scale. We
showed~\cite{CHN} that for $m_{\widetilde{sb}_{1R}} \sim 200$ GeV,
$m_{\tilde g} \sim 500$ GeV and $\sigma \sim \pi/4-\pi/2$, the
model could account for the $\phi K_S$ $CP$ violation anomaly,
with a host of predictions. Basically, in formulas analogous to
Eq.~(\ref{eq:Amp}), the primed terms are generated.
We observed~\cite{ACH} that the dominant effect is in
$c_{12}^\prime$ (and $c_{7,11}^\prime$), all other terms are
small. Thus, it seems natural to consider the impact of this NP
model on $\phi K^*$ and $\omega K^*$.

With the above parameter values and $\sigma = 90^\circ$, we plot
the branching fractions of $B\to \phi K^*$ and $\omega K^*$ vs
$\kappa$ in Figs.~3(a) and 3(b), respectively.
From Fig. 3(a) we see that $\kappa < 0$ is no longer viable since,
contrary to experiment, the parallel rate becomes too large
compared with the perpendicular one. We take $\sigma = 90^\circ$
because otherwise the parallel rate rises too slowly for $\kappa >
0$.
But $\kappa \sim +0.25$ can be viewed as a solution of the $\phi
K^*$ polarization anomaly in this NP model.
With $\kappa$ fixed this way, the $\omega K^*$ results of
Fig.~3(b) are predictions, and previous remarks continue to apply.
Note that the parallel component may be the largest.

%
%
%
\begin{figure}[t!]
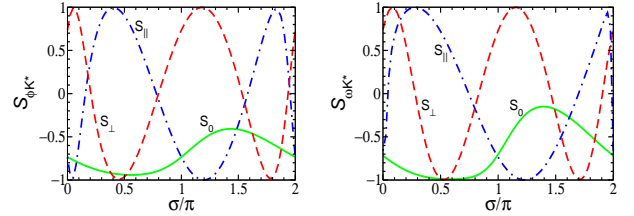

\smallskip  
\includegraphics[width=1.5in,height=1.1in,angle=0]{SphiKstar}
\hspace{2mm}
\includegraphics[width=1.5in,height=1.1in,angle=0]{SomegaKstar}
\smallskip\smallskip
\caption{ ${\cal S}_\lambda$ vs $CP$ phase $\sigma$ for (a) $\phi
K^{*0}$ and (b) $\omega K^{*0}$ for $\lambda = 0$ (solid),
$\parallel$ (dotdash) and $\perp$ (dash), with $\kappa = +0.25$. }
\label{fig:Slambda}
\end{figure}

The interest in discussing NP is not so much about the
polarization anomaly itself, but to predict associated $CP$
violating asymmetries. Since the $B_d$ lifetime difference is
negligible, and since we do not consider rescattering and
associated strong phases, the main observable is the mixing
dependent $CP$ asymmetry for $\bar B^0\to \phi K^{*0}$ and $\omega
\bar K^{*0}$ in each polarization component,
\begin{eqnarray}
S_\lambda= \frac{2\xi\;{\rm Im}(\frac{q}{p}\;\overline {\cal
A}_\lambda {\cal A}_\lambda^\ast)} {\overline {\cal
A}_\lambda^2+{\cal A}_\lambda^2},
 \label{eq:Slambda}
\end{eqnarray}
which is the coefficient of the $\sin\Delta m\Delta t$ oscillation
term, $q/p$ contains the $B^0$ mixing phase $\sin2\phi_1$, and
$\overline {\cal A}_\lambda$ is the amplitude of the conjugate
process.
Taking $\kappa = 0.25$ as example, we plot ${\cal S}_\lambda$ vs
$CP$ phase $\sigma$ in Figs.~4(a) and 4(b) for $\phi K^{*0}$ and
$\omega K^{*0}$, respectively.

The $\sigma$ dependence can be understood as follows. Let us write
a particular polarization amplitude as ${\cal A} = a\,
e^{i\theta_1} e^{i\delta_1} + b\, e^{i\theta_2} e^{i\delta_2}$,
where the first term is the SM contribution sans chromodipole
penguin (hence $CP$ phase $\theta_1 \cong 0$), while the second
term is $\propto c_{12}\mp c_{12}^\prime$ (hence $\theta_2$
depends on $\sigma$). Assuming that strong $\delta$ phases are
small, and taking the $\phi K^{*0}$ case as example, it is easy to
see that for $S_0$, one has $a^2 \gg b^2$. The leading effect is
then the SM $CP$ phase in $B_d$ mixing (modulo the $CP$ eigenvalue
$\xi$), plus a simple ($\sim \sin\sigma$) modulation around this
constant value.
For $S_\parallel$ and $S_\perp$, however, one has $a^2 \ll b^2$
because of $1/m_b$ suppression of usual terms. Thus,
$S_{\parallel}$ and $S_\perp$ is determined by $(c_{12} \mp
c_{12}^\prime)^2$ (see Eq.~(\ref{eq:Amp})) hence close to
$\sin2\sigma$ variation from the SM expectation value of
$\mp\sin2\phi_1 \simeq \mp 0.74$.
Note that for $\sigma \simeq \pi/2$, one would find
$S_{\parallel}$ and $S_\perp$ to be of similar strength to
expectation, {\it but with opposite sign}, which is analogous to
$S_{\phi K_S}$~\cite{CHN}. Unfortunately one is not free from
hadronic uncertainties.
It is important to separate the parallel and perpendicular
components to make such measurements, for otherwise they would
dilute each other out.

%
%
%
\begin{figure}[t!]
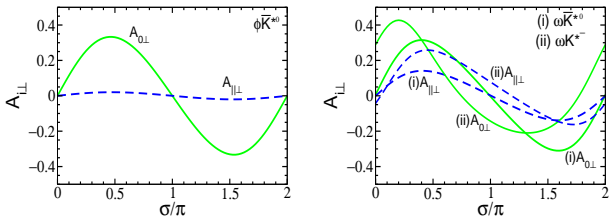

\smallskip  
\includegraphics[width=1.5in,height=1.1in,angle=0]{Lam_phiKstar}
\hspace{2mm}
\includegraphics[width=1.5in,height=1.1in,angle=0]{Lam_omegaKstar}
\smallskip\smallskip
\caption{ $T$ violation parameter $A_{i\perp}$ vs $\sigma$ for
$B\to$ (a) $\phi K^*$ and (b) $\omega K^*$ for $i = 0$ (solid),
$\parallel$ (dash) components. } \label{fig:Todd}
\end{figure}

Another intriguing measure, which needs neither oscillation
measurement nor tagging, is the triple product $T$-violation
parameter~\cite{DLSS},
\begin{eqnarray}
A_{i\perp} = \frac{1}{2}\, \Biggl(
   \frac{{\rm Im}\,({\cal A}_\perp {\cal A}^\ast_i)}
        {\sum|{\cal A}_\lambda|^2}
 + \frac{{\rm Im}\,({\cal\overline A}_\perp {\cal\overline A}^\ast_i)}
        {\sum|{\cal\overline  A}_\lambda|^2}
 \Biggr),
\end{eqnarray}
from interference pattern in angular analysis.
We plot $A_{i\perp}$ vs $\sigma$ in Figs.~5(a) and 5(b) for $B\to
\phi K^*$ and $\omega K^*$, respectively. For $\phi K^*$, there is
little difference between charged and neutral mode. But for
$\omega K^*$, the tree contribution brings in the SM $CP$ phase
$\phi_3$, and there is some difference between charged vs neutral
modes.
We note with interest that BaBar has measured~\cite{BaBar04}
$A_{0\perp} = 0.11\pm0.05\pm0.01$ and $A_{\parallel\perp} =
-0.02\pm0.04\pm0.01$ for $B\to \phi K^*$, which is consistent with
our results. The agreement is not yet very significant. However,
given the error bars, significant results may soon become
available with improved data sets.

Some remarks are in order.
The $b\to sg$ parton level process, with non-negligible rate at
few $\times 10^{-3}$ level~\cite{Hou88} (reaching $10^{-2}$ in the
NP case with $\sigma = 90^\circ$), has so far been rather elusive
for direct access. But given that it is the QCD analog of $b\to
s\gamma$, it is certainly quite important. It would be amusing if
the polarization anomaly would turn out to be the harbinger of
$b\to s$ penguin involving on-shell gluon emission.
Second, the recoil system against $\phi_T$ is in general
complicated. One could consider searching for $\phi_T/\omega_T +
(Kn\pi)_V$, or even a recoil tensor meson.
Conversely, the transverse gluon could fragment into any flavor
singlet low-mass hadronic system that has total spin 1. One may
therefore wish to search for $B\to$ ``$V$"$K^*_T$ where ``$V$"
stands for some low-mass singlet vector configuration.
Third, it would be nice if the on-shell $b\to sg$ penguin could
help resolve the problem of large $B\to \eta^\prime K$ rate and
the {\it finite} $B\to \omega K$ rate. But it is unclear how the
mechanism we outlined could feed longitudinal vector meson
production. Perhaps the polarization can be left behind in form of
hadronic excitation as the gluon ``energy packet" leaves the $B$
meson carcass.
Finally, we note that inclusive $b\bar q\to q_1 \bar q_2$
annihilation rate is slightly larger~\cite{Hou88} than $b\to sg$,
which has been invoked~\cite{Kagan} for generating the transverse
components of $B\to \phi K^*$. However, the proposal is not
predictive and needs additional arguments for $B\to \rho K^*$.
Likewise, the $D_s^{(*)}\bar D^{(*)}$ rescattering
picture~\cite{Colangelo} would also have trouble with $\rho K^*$.

In conclusion, we have given an {\it ad hoc} but simple one
parameter model where transverse components for $B\to \phi K^*$
descend from on-shell $b\to sg$, where $\phi_T$ is a transverse
gluon fragment. Similar behavior is predicted for $B\to \omega
K^*$ but not $\rho K^*$ and $\rho\rho$ modes.
Although the picture is generic, New Physics $CP$ phases could
generate opposite sign $CP$ violation in $B\to \phi K^*$, $\omega
K^*$ transverse components, as well as $T$-violating triple
product observables.

\vskip 0.3cm \noindent{\bf Acknowledgement}.\ \ We thank
K.F.~Chen, A.~Kagan, H.n.~Li, R. Sinha and A.~Soddu for
discussions. This work is supported in part by grants
NSC-93-2112-M-002-020 and NSC-93-2811-M-002-053, and NCTS/TPE.


\begin{thebibliography}{99}
%
\bibitem{PDG}
S. Eidelman {\it et al.} (Particle Data Group), Phys. Lett. B {\bf
592}, 1 (2004).
%
\bibitem{HFAG}
For updated results, see Heavy Flavor Averaging Group [HFAG]:
http://www.slac.stanford.edu/xorg/hfag/.
%
\bibitem{Belle03}
K.F. Chen, A. Bozek {\it et al.} [Belle Collab.], Phys. Rev. Lett.
{\bf 91}, 201801 (2003).
%
\bibitem{BaBar03}
B.~Aubert {\it et al.}  [BaBar Collab.], Phys.\ Rev.\ Lett.\ {\bf
91}, 171802 (2003).
%
\bibitem{BaBar04}
%
B.~Aubert {\it et al.}  [BaBar Collab.], hep-ex/0408017.
\bibitem{Kagan}
A.L. Kagan, hep-ph/0405134 and hep-ph/0407076.
%
\bibitem{phiKsBelle}
K.~Abe {\it et al.}  [Belle Collab.], Phys.\ Rev.\ D {\bf 67},
031102 (2003);
Phys.\ Rev.\ Lett.\  {\bf 91}, 261602 (2003).
%
\bibitem{phiK0BaBar}
B.~Aubert {\it et al.}  [BaBar Collab.],
hep-ex/0403026.
%
\bibitem{Hou88}
W.S. Hou, Nucl. Phys. B {\bf 308}, 561 (1988).
%
\bibitem{rrBelle}
J. Zhang, M. Nakao {\it et al.} [Belle Collab.], Phys.\ Rev.\
Lett.\  {\bf 91}, 221801 (2003).
%
\bibitem{rrBaBar}
B.~Aubert {\it et al.}  [BaBar Collab.], Phys.\ Rev.\ D {\bf 69},
031102 (2004).
%
\bibitem{LCSR}
P. Ball and V.M. Braun, Phys. Rev. D {\bf 58}, 094016 (1998); P.
Ball, hep-ph/0306251. 
%
%
\bibitem{CHN}
C.K. Chua, W.S. Hou and M. Nagashima, Phys. Rev. Lett. {\bf 92},
201803 (2004). 
%
\bibitem{ACH}
C.K. Chua and W.S. Hou, Phys. Rev. Lett. {\bf 86}, 2728 (2001);
%
A. Arhrib, C.K. Chua and W.S. Hou, Phys.\ Rev.\ D {\bf 65}, 017701
(2002).
%
\bibitem{DLSS}
A. Datta and D. London, 
Int.\ J.\ Mod.\ Phys.\ A {\bf 19}, 2505 (2004);
%
D. London, N. Sinha and R. Sinha, hep-ph/0304230.
%
\bibitem{Colangelo}
P. Colangelo, F. De Fazio and T.N. Pham, hep-ph/0406162.
%
%
%
%
\end{thebibliography}
\end{document}